\documentstyle[epsf,rotate]{aipproc}
\newcommand{\gsim}{\raisebox{-0.7ex}{$\stackrel{\textstyle >}{\sim}$ }}

\begin{document}
\title{Effective Field Theory in \\
Nuclear Physics\footnote{
Talk presented at the 
{\it 7th Conference on the Intersections of 
Particle and Nuclear Physics},
Quebec City, Canada,  May 22-28, 2000.
NT@UW-00-018.} 
}

\author{Martin J. Savage}
\address{Department of Physics, University of Washington, \\
Seattle, WA 98915 \\ and \\
Jefferson Lab., 12000 Jefferson Avenue, Newport News, \\
Virginia 23606.}

\maketitle

\begin{abstract}
I review recent developments in the 
application of effective field theory to nuclear physics.
Emphasis is placed on precision two-body calculations 
and efforts to formulate the nuclear shell model in terms of
an effective field theory.
\end{abstract}

\section*{Introduction}

A question that I have been asked many times
is 
``{\it Why use Effective Field Theory in Nuclear Physics}''?
The simple and somewhat glib answer to this question 
is that the only other option
that one has to using an effective field theory (EFT) is to use the 
``{\it Theory of Everything}'' (TOE), 
string theory or some derivative thereof.
All other descriptions {\bf must} be incomplete at some level
and when precise predictions are compared with precise measurements,
differences will become obvious.
It is a daunting prospect for us (maybe only me) to 
use the TOE to compute low-energy hadronic processes, and
in fact, it is quite silly to even consider such calculations.
After all, we know that processes in QED 
can be computed to high precision
which agree with experimental observations, without knowing anything about 
physics at the Planck scale, $M_{\rm pl}$.

The renormalizability of QED assures that ultra-violet divergences, arising
from our lack of understanding of physics at short-distances, 
can be explicitly removed by a few constants (electric charge and fermion
mass),
allowing observables to be related to each other to arbitrary precision.
In contrast, EFT's are non-renormalizable but are still
predictive when a systematic
power counting in small expansion parameters can be established.
Relations between observables at a given precision will involve a finite 
number  of constants that are not dictated by
the symmetries of the EFT  alone.
For instance, the standard model is a renormalizable field theory
but processes at energies much less than the scale
of electroweak symmetry breaking can be described by an
non-renormalizable EFT of reduced symmetry,
$SU(3)_c\otimes SU(2)_L\otimes U_Y(1) 
\rightarrow SU(3)_c\otimes U_{\rm em}(1)$,
where weak interactions are incorporated by higher-dimension
four-fermi operators.
Even with quantum effects, the theory provides a systematic expansion of
observables in terms of $q^2/ M_W^2$ and $m_f^2/M_W^2$, where $m_f$ is a
fermion mass, $M_W$ is the mass of the weak gauge bosons, and $q$ is the
external momentum.

\begin{figure}[!ht] 
\epsfysize=2.5in
\centerline{\epsfbox{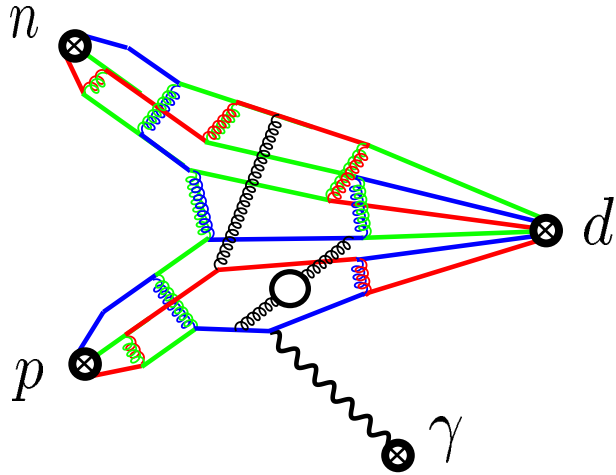}}
\vspace{10pt}
\caption{A ``cartoon'' of a contribution to  $np\rightarrow d\gamma$
in QCD.}
\label{fig1}
\end{figure}

If one were interested in calculating the cross section for $np\rightarrow
d\gamma$--radiative neutron capture by a proton to form a 
deuteron--directly from QCD,
then a lattice calculation is the only technique available.
The lattice calculation
will provide an unambiguous cross section in terms of the 
quark masses and $\Lambda_{\rm QCD}$, or equivalently in terms of other
hadronic observables.
A cartoon of a contribution to $np\rightarrow d\gamma$
in terms of perturbative quarks and gluons 
is shown in Figure~\ref{fig1}.
Sources of quarks and gluons that have non-zero overlap with the proton,
neutron, deuteron and a source for the photon would used to generate the
amplitude for $np\rightarrow d\gamma$.
It is clear that a significant amount of work goes into forming the hadronic
states themselves, let alone computing the interaction terms.
In addition, as the deuteron has such a small binding energy and hence is quite
extended compared to the nucleon, considerable effort will be required to
generate the deuteron itself.
Unfortunately, at this point in time the lattice community is not even 
close to being able to perform this multi-hadron calculation.
Indeed, the deuteron itself 
remains to be generated in lattice 
calculations.~\footnote{During the {\it Effective Field Theory}
workshop to be held at the {\it Institute for Nuclear Physics}
at the {\it University of Washington} during the summer of 2000,
efforts will be made to estimate the computer resources 
necessary to determine the deuteron binding 
energy  from lattice QCD~\cite{Lepage}.}
If all nuclear lengths scales were of order the
chiral symmetry breaking scale $\Lambda_\chi$ then (very) naively
lattice computations of nuclear observables would not be that much harder
than computations in the single-nucleon sector.
However, there are several low-energy length scales that play important roles
in nuclear physics.
Firstly $\Lambda_\chi$, below which a hadronic description makes sense,
and higher-dimension operators are induced that describe 
contributions from scales above $\Lambda_\chi$.
Secondly, the scale of the repulsive part of the nucleon-nucleon interaction,
which is conventionally modeled by the exchange of vector mesons 
(far from mass-shell) and numerically is of order $\Lambda_\chi$.
Thirdly, the mass of the pion, which is much less than $\Lambda_\chi$ due
to its special status as a pseudo-Goldstone boson.
Finally, nuclear binding energies which are much smaller than one would 
naively guess.

\begin{figure}[h!] 
\epsfysize=3.2in
\centerline{\epsfbox{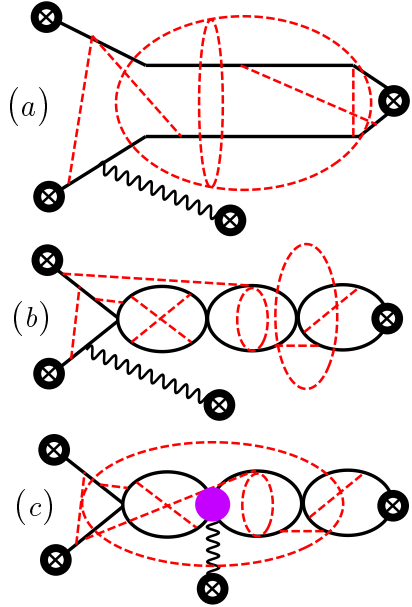}}
\vspace{10pt}
\caption{Contributions to 
$np\rightarrow d\gamma$ in a theory with nucleons, pions and
photons.
Diagram~(a) shows a contribution from $\pi$-exchange alone,
while
diagram~(b) shows a contribution from $\pi$-exchange and from
short-distance interactions.
Diagram~(c) shows a contribution from a local, gauge invariant operator
not constrained by nucleon-nucleon scattering data.
}
\label{fig2}
\end{figure}

If one is interested in this process at energies much less than
$\Lambda_\chi$, 
or the mass of the $\rho$-meson, 
but comparable to the mass of the pion,
then it should be sufficient to use an
EFT with only nucleons, pions and photons as dynamical degrees of freedom.
All contributions from higher mass scales will be encapsulated 
in the infinite number of higher dimension operators that arise in the 
momentum and chiral expansions.
Some diagrams that will contribute to $np\rightarrow d\gamma$ are shown 
in  Figure~\ref{fig2}.
Unlike most EFT's that one encounters, higher dimension operators 
(dim-6)
involving the nucleon field play a central 
role\cite{Weinberg}-\cite{KSW}.
The fact that the deuteron is barely bound,
with a binding energy of $B=2.2~{\rm MeV}$,
requires a fine-tuning between pion exchange and short-distance physics.
Naively, one would expect a binding energy set by $f_\pi$, the pion decay
constant, much larger than one finds in nature.
Therefore, the class of diagrams shown in 
Figure~\ref{fig2}(b) is not expected to be suppressed compared to those in 
Figure~\ref{fig2}(a).
In addition, contributions from diagrams shown in Figure~\ref{fig2}(c)
must be included.
These arise from an insertion of operators that are gauge invariant by
themselves, and are not related in any way to operators describing 
nucleon-nucleon scattering.
They arise from short-distance physics and have a scale typically set by 
$\Lambda_\chi$.

\begin{figure}[h!] 
\epsfysize=2.5in
\centerline{\epsfbox{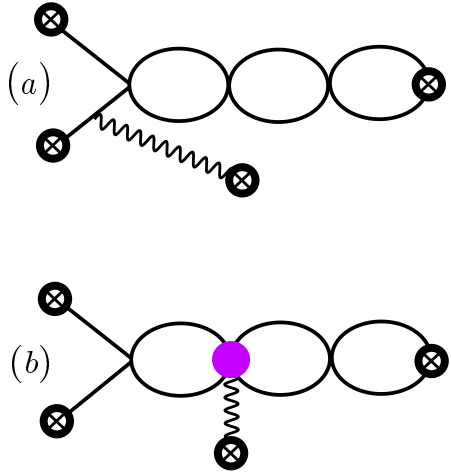}}
\vspace{10pt}
\caption{Diagrams that  contribute to 
$np\rightarrow d\gamma$ in a theory with nucleons and photons.
Diagram~(a) shows a contribution from 
interactions between nucleons.
Diagram~(b) shows a contribution from a local, gauge invariant operator
that does not contribute (at tree-level) to
nucleon-nucleon scattering.
}
\label{fig3}
\end{figure}

Continuing our descent in energy,
if one is interested in this process at energies much less than 
the pion mass, $m_\pi$, 
then it should be sufficient to use an
EFT with only nucleons and photons as dynamical degrees of freedom.
All contributions from mass scales greater than $m_\pi$ will be encapsulated 
in the infinite number of higher dimension operators that arise in the 
momentum expansion (chiral symmetry is explicitly broken, leaving isospin
symmetry as the only relic of the flavor symmetries).
Some diagrams that will contribute to $np\rightarrow d\gamma$ are shown 
in  Figure~\ref{fig3}.
It is important to realize that nucleon-nucleon scattering described by this
EFT  uniquely reproduces Effective Range Theory (ERT)\cite{ERtheory}.
However, for all other observables, such as those involving electroweak gauge
fields, ERT 
(e.g. \cite{ERtheory,Noyes})
is seen to be an uncontrolled approximation to EFT (e.g. \cite{CRSa}).

In the following sections I attempt to indicate the status
of EFT descriptions in the various energy regimes.
Firstly, I will discuss 
low-energy $|{\bf p}|\ll m_\pi$ processes involving two and three
nucleons, focusing on the recent high precision
calculations that have been performed. 
Secondly, the issues, results  and the present roadblocks to successfully
describing the intermediate energy regime  $|{\bf p}|\gsim m_\pi$
are presented.
Finally, efforts to translate the 
understanding gained in EFT developments to many-nucleon systems
are described.
Such translation is necessary in order to achieve the ultimate goal of 
having a perturbative theory of nuclei that faithfully
reproduces QCD.

\section*{$np\rightarrow d\gamma$ at Low Energies}

During the past year there has been considerable focus placed on the 
radiative capture process $np\rightarrow d\gamma$.
Firstly, 
it was pointed out\cite{BNTTa} that the uncertainty in the cross section
for $np\rightarrow d\gamma$ contributed  significantly  to the 
uncertainties in the predictions of Big-Bang-Nucleosynthesis (BBN)
of light element abundances.
This resulted from the lack of  
data in the 
energy region important for BBN from either $np\rightarrow d\gamma$ or
$\gamma d\rightarrow np$.
Further, available 
potential model calculations of these
processes\cite{ENDF} 
are undocumented and error estimates are absent.
Tools have recently been developed that allow for a $\sim 1\%$
calculation of the cross section with EFT\cite{CSa,Ra}.
This was facilitated in part by realizing that
it is advantageous
to get not only the location of the deuteron pole correct, 
but  also the normalization of the deuteron s-state component.
This had long been implemented in the methods of \cite{Parka,Parkeft}, and 
implicit in the construction of Weinberg\cite{Weinberg,Bira}, 
but was only recently implemented in the dimensionally regulated 
EFT\cite{PRSa}.
Finally, there are experimental efforts to measure the small 
isoscalar $E2_S$ and $M1_S$ 
amplitudes contributing to $np\rightarrow d\gamma$ using polarized neutrons on
polarized protons\cite{nppolexpt}.
At the second workshop on {Effective Field Theory in Nuclear Physics}
held at the {\it Institute for Nuclear Physics} at the 
{\it University of Washington} in 1999, 
Mannque Rho challenged the 
participants to compute the $E2_S$ and $M1_S$ amplitudes
with the group whose predictions are verified
experimentally winning a bottle of exceptional wine.
With such a wonderful prize at stake, many workshop participants redirected
their efforts to this project.
This is now known as the {\it Rho-challenge}.

\subsection*{$np\rightarrow d\gamma$ for Big-Bang Nucleosynthesis
\footnote{I thank Gautam Rupak for allowing me to present his 
results in this section}}

\begin{figure}[h!] 
\epsfysize=2.0in
\centerline{\epsfbox{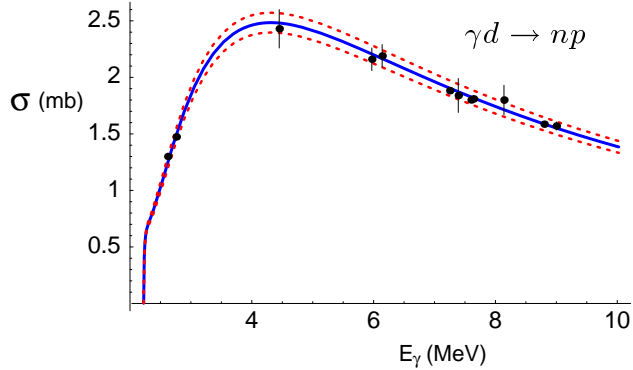}}
\vspace{10pt}
\caption{
The photodissociation cross section for
$\gamma d\rightarrow n p$.
The solid line results from
eqs.(\ref{unpol}) and (\ref{eq:EMamps}) 
\protect\cite{CSa} with
$L_{np}$ determined by the cross section
for cold $np\rightarrow d\gamma$.
The dashed lines denote the theoretical uncertainty.
Rupak has further 
reduced this uncertainty to below $1\%$\protect\cite{Ra}.
}
\label{fig4}
\end{figure}

As existing potential model calculations of $np\rightarrow d\gamma$
are undocumented and error estimates unavailable,
a $5\%$ uncertainty was assigned to the cross section as input into
BBN codes\cite{SKMa}.
It would be somewhat dismal if, 
after several decades of investigations into nuclear physics, 
the cross section for this process was uncertain at the 
$5\%$ level.
However, EFT calculations have demonstrated that
the actual uncertainty is much less than $5\%$.
EFT is a well defined method of calculation 
and estimates of the uncertainty in a given calculation can be
made by considering the magnitude of 
higher order terms that have been omitted.
The expression for the cross section for $np\rightarrow d\gamma$
valid at the $\sim 3\%$ level (with  nonrelativistic kinematics)
is
\begin{eqnarray}
 \sigma & = & {4\pi\alpha \left(\gamma^2+|{\bf P}|^2\right)^3
\over \gamma^3 M_N^4 |{\bf P}|}
\left[\  |\tilde{X}_{M1}|^{2}
  \ +\ |\tilde{X}_{E1}|^{2}\ 
\right]
\ \ \ .
\label{unpol}
\end{eqnarray}
The isovector $M1_V$ and $E1_V$ amplitudes are
\begin{eqnarray}
 |\tilde{X}_{M1}|^2 & = & 
{\kappa_1^2 \gamma^4 \left( {1\over a_1}-\gamma\right)^2 \over
\left({1\over a_1^2} +|{\bf P}|^2\right) \left(\gamma^2 + |{\bf P}|^2 \right)^2
}
\left[Z_d
- r_0 { \left( {\gamma\over a_1}+|{\bf P}|^2\right)  |{\bf P}|^2 \over
\left({1\over a_1^2} +|{\bf P}|^2\right) \left( {1\over a_1}-\gamma\right)}
- {L_{np}\over\kappa_1}{M_N\over 2\pi}
{ \gamma^2 + |{\bf P}|^2 \over {1\over a_1}-\gamma}
\right]
\nonumber\\
 |\tilde{X}_{E1}|^2 & = & 
{|{\bf P}|^2 M_N^2 \gamma^4 \over \left(\gamma^2+|{\bf P}|^2\right)^4}
\left[ Z_d \ +\ {M_N\gamma\over 6\pi}
\left({\gamma^2\over 3}+|{\bf P}|^2\right)\overline{C}^{(P)}
\right]
\ \ \ ,
\label{eq:EMamps}
\end{eqnarray}
where $\kappa_1$ is the isovector nucleon magnetic moment,
$a_1=-23.714\pm 0.013~{\rm fm}$ 
is the scattering length in the ${{}^1\kern-.14em S_0}$ channel,
$r_0=2.73\pm 0.03~{\rm fm}$ is the effective range in the 
${{}^1\kern-.14em S_0}$ channel,
$\gamma=\sqrt{M_N B}$ is the deuteron binding momentum,
$\overline{C}^{(P)}$ is a number derivable from the 
nucleon-nucleon p-wave amplitudes, 
$Z_d~=~1/(1~-~\gamma~\rho_d)$ 
is the residue 
of the ${{}^3\kern-.14em S_1}$ nucleon-nucleon amplitude at the 
deuteron pole, and
$\rho_d$ is the effective range in the  ${{}^3\kern-.14em S_1}$ channel.
There is also a contribution from an operator that is not related by gauge 
invariance to nucleon-nucleon scattering, $L_{np}$.
This is the coefficient of a local operator involving four-nucleon fields and a
magnetic photon.
For incident neutrons  with speed $|v|=2200~{\rm m/s}$ the cross section
for capture by protons at rest is measured to be 
$\sigma^{\rm expt}_{\rm cold} = 334.2\pm 0.5~{\rm mb}$\cite{CWCa}.
The value of  $L_{np}$ is fixed by 
requiring that the  
expressions in eqs.~(\ref{unpol}) and (\ref{eq:EMamps})
reproduce $\sigma^{\rm expt}_{\rm cold}$.
The amplitudes in eq.~(\ref{eq:EMamps}) have been computed to 
one higher order
by Rupak\cite{Ra}, including relativistic effects and 
the appearance of an E1 counterterm,
providing a $\sim 1\%$ calculation of $np\rightarrow d\gamma$.
Once $L_{np}$ has been fixed, the photo-dissociation cross section
$\gamma d\rightarrow n p$ can be determined,
and is shown in Figure~\ref{fig4}.
One finds that this relatively simple analytic expression reproduces
the data very well, once the counterterm $L_{np}$ has been determined at 
a given energy.
An idea of the convergence of the effective field theory calculation
can be obtained from the numerical results of Rupak\cite{Ra},
\begin{eqnarray}
\sigma (2~{\rm MeV}) & = & 
0.0218 \left( 1 + 0.6389 + 0.0135 - 0.0053 - 0.0001 + ...\right)~{\rm fm^2}
\nonumber\\
\sigma (20~{\rm keV}) & = & 
0.1917 \left( 1 + 0.1076 + 0.0001 + ...\right)~{\rm fm^2}
\ \ \ ,
\label{eq:conv}
\end{eqnarray}
which are both seen to converge rapidly.
\begin{table}[!h]
\caption{
$\sigma (np\rightarrow d\gamma)$ as a function of the 
nucleon center-of-mass energy, $E$.
The asterisk denotes an input.
}
\begin{tabular}[h]{ccc}
$E$ (MeV) & total EFT $\sigma$ (mb)\cite{Ra} 
& ENDF\ $\sigma$ (mb)\cite{ENDF}  \\ \hline
$1.264\times 10^{-8}$ & $334.2^{\ (*)}$ & 332.0\\
$5.0\times 10^{-4}$ & 1.668(0) & 1.660\\
$1.0\times 10^{-3}$ & 1.172(0) & 1.193\\
$5.0\times 10^{-3}$ & 0.4982(0) & 0.496\\
$1.0\times 10^{-2}$ & 0.3324(0) & 0.324\\
$5.0\times 10^{-2}$ & 0.1081(0) & 0.108\\
0.100 & 0.06352(0) & 0.0633\\
0.500 & 0.0341(1) & 0.0345\\
1.00 & 0.0349(4) & 0.0342
\end{tabular}
\label{table1}
\end{table}
The cross section at various energies computed with EFT by Rupak\cite{Ra},
along with those from the on-line nuclear data center\cite{ENDF} are shown in 
Table~\ref{table1}.
As expected the EFT calculation agrees with the 
numerical values from
\cite{ENDF} at the $\sim 1\%$ level.

\subsection*{Isoscalar M1 and E2  Amplitudes in $np\rightarrow d\gamma$}

As mentioned earlier, the {\it Rho Challenge} focused 
on the $M1_S$ and $E2_S$ isoscalar amplitudes
that contribute to $np\rightarrow d\gamma$, so that predictions 
can be compared with the imminent measurement of polarization 
observables\cite{nppolexpt}.
Two works were completed soon after the challenge was issued, one by 
Kubodera, Park, Min, and Rho\cite{PKMRnp}, 
and one by Chen, Rupak and myself\cite{CRSb}.

There are a couple of angular distributions that can be measured in
$\vec{n}+\vec{p}\rightarrow d\gamma$, but if in addition, the polarization of
the $\gamma$ can be measured one finds that
there is a different cross section for production of right-handed
versus left-handed
circularly polarized photons.
Defining the asymmetry $A^\gamma (\theta)$ to be the 
ratio of the  difference to the sum of these
cross sections, 
\begin{eqnarray}
  A^\gamma_{\eta_{\rm n}} (\theta) & = &
  \eta_{\rm n} \left[\ 
    \left( P_\gamma (M1)\ +\   P_\gamma (E2) \right)\ \cos\theta
    \ +\   P_\gamma (E1)\ \sin^2\theta
    \right]
\ \ \ ,
\label{eq:gamas}
\end{eqnarray}
where the $P_\gamma (\Pi L)$ are combinations of the 
$M1_V$, $M1_S$, $E1_V$ and $E2_S$ amplitudes,
and $\eta_{\rm n}$ is the neutron polarization vector.

The amplitudes and polarizations are computed in two very different ways.
In \cite{PKMRnp} EFT wavefunctions are developed with a coordinate-space
cut-off, which are then used to determine matrix elements of the various
electric and magnetic multipole operators.  
Pions appear as dynamical degrees of freedom and determine the
long-range
part of the nucleon-nucleon interaction.
In addition, counterterms are
included via short-distance interactions (e.g. ``delta-shell'' and others)
so  that the magnetic and quadrupole moment of the deuteron are
recovered.
A very different construction is used in \cite{CRSb}.
The EFT without pions is used and divergences are
dimensionally regulated.  As in \cite{PKMRnp}, the 
four-nucleon-one photon counterterms
are chosen to recover the deuteron magnetic and quadrupole moments,
to give
$P_\gamma (M1)\ =\  -7.1\times 10^{-4}$, 
$P_\gamma (E2) \  = \  -3.5\times 10^{-4}$
and a total of
$P_\gamma =-1.06\times 10^{-3}$ in the forward direction, 
approximately $2/3$ of the
experimentally determined value of\cite{Bazh}
$P_\gamma^{\rm expt} =-(1.5\pm 0.3) \times 10^{-3}$.
Given the large uncertainty in the calculation of the $M1_S$ amplitude,
and the uncertainty of the measurement, the two are not inconsistent.
$P_\gamma (M1)= -7.1\times 10^{-4}$ agrees
with the results of Burichenko and Kriplovich\cite{BKa}
of $P_\gamma (M1)= -7.0\times 10^{-4}$ from a Reid soft-core calculation,
but is somewhat less than their zero-range calculation of
$P_\gamma (M1)= -9.2\times 10^{-4}$.
However, given the large uncertainty in the $M1_S$ amplitude
of \cite{CRSb}, both values are
consistent.
$P_\gamma (E2)  =  -3.5\times 10^{-4}$ calculated in
\cite{CRSb} agrees well with that computed in
\cite{PKMRnp},
and therefore these observables do not distinguish between the two 
EFT methods.

\section*{Weak Interactions of the Deuteron}

Weak interaction processes involving the deuteron are central
to current research efforts in nuclear physics.
In addition to the accelerator based programs to elucidate the 
flavor structure of the nucleon, such as the SAMPLE experiments\cite{SAMPLE} 
at Bates,
the interactions between neutrinos and the deuteron 
form the core of our efforts to learn  about the neutrino and 
look beyond the standard model of electroweak interactions.
Both in production, e.g. $pp\rightarrow d e^+\nu_e$, and in detection
at SNO (Sudbury Neutrino Observatory), 
e.g. $\nu_\mu d\rightarrow \nu_\mu np$, charged and neutral current 
weak interaction matrix elements 
between the deuteron and continuum states are required.

Much effort over the past few decades has been put into calculating the
production mechanism $pp\rightarrow d e^+\nu_e$, both from  standard
non-relativistic quantum mechanics\cite{PotBah}, 
and from  sophisticated  potential model techniques\cite{PotSch}.
Recently, EFT has been applied to this process
by Kong and Ravndal\cite{KRpp}
and by Park, Kubodera, Min and Rho\cite{PKMRpp}
giving elegant expressions and 
numerical values of the weak capture cross section that are consistent
with previous estimates\footnote{
There has also been recent work in \cite{Ivan} that I have so far failed
to comprehend.}. 
The cross section depends somewhat on 
the value of a 
four-nucleon-one-weak-gauge-boson interaction, with coefficient
$L_{1,A}$,  as defined in~\cite{BCa}.

\begin{figure}[!t]
\epsfysize=2.0in
\centerline{\epsfbox{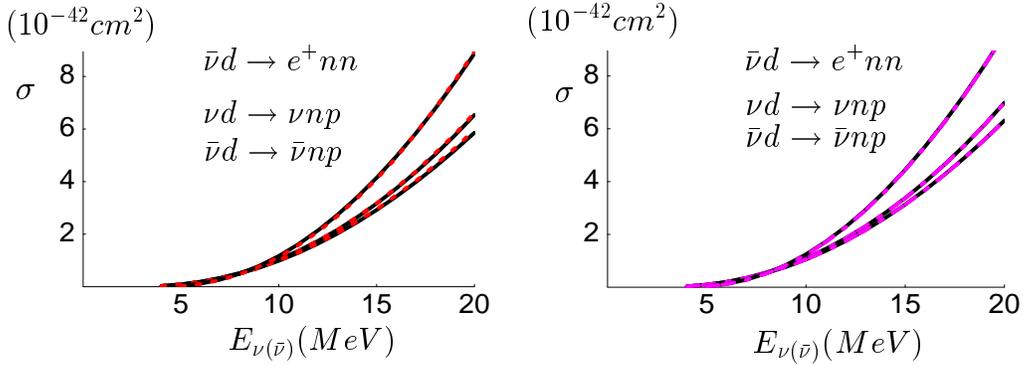}}
\caption{ 
Inelastic $\protect\nu (\bar{\protect\nu})d$ 
cross-sections versus incident 
$\protect\nu (\bar{\protect\nu})$ energy. 
The solid curves in the left graph are KN results\protect\cite{KN} 
while the dot-dashed curves,
which lie on top of the solid curves, are NLO in EFT with 
$L_{1,A}=6.3\ {\rm fm}^{3}$. 
The solid curves in the right graph are YHH results\protect\cite{YHH} 
while the dashed curves, which also lie  on top of the solid curves,
are NLO in EFT with $L_{1,A}=1.0\ {\rm fm}^{3}$.  }
\label{fig5}
\end{figure}

The detection reactions, $\nu d\rightarrow n p\nu$,
$\nu_e d\rightarrow e^- pp$ and $\overline{\nu}_e d\rightarrow e^+ nn$
had been looked at by two groups,
Ying, Haxton and Henley (YHH)\cite{YHH}
and Kubodera and Nozawa (KN)\cite{KN},
using sophisticated potential
models.
The two sets of calculations
differ at the $5\%$ level, due to different treatments 
of  meson-exchange  currents (MEC).
Recently, Butler and Chen\cite{BCa} have determined the break-up
cross sections with EFT.
Only  $L_{1,A}$ needs to be fixed in order
to perform a $\sim 1\%$ calculation.
The YYH and KN numerical results can be
recovered with different choices of $L_{1,A}$,
as shown in Figure~\ref{fig5}\footnote{
I thank Malcolm Butler and Jiunn-Wei Chen for allowing me to reproduce their
figures}, 
confirming 
that the difference between the two potential-model calculations 
is short-distance in origin.
Therefore, to predict the break-up cross section with a precision of better
than $\sim 5\%$, one has two options.
Firstly, compute the $\beta$-decay of tritium, and use this to determine
the counterterm $L_{1,A}$ in the EFT, or equivalently
the MEC's in the potential models\footnote{
This method of fixing the MEC's has already been
implemented for $pp\rightarrow d e^+\nu_e$\cite{PotSch}.
}.
Secondly, one can perform an experiment to measure one of the break-up cross
sections to high accuracy, and thereby extract $L_{1,A}$, or the MEC's.
Such an experiment is currently under consideration\cite{Avig}.
\begin{figure}[!t]
\epsfysize=2.0in
\centerline{\epsfbox{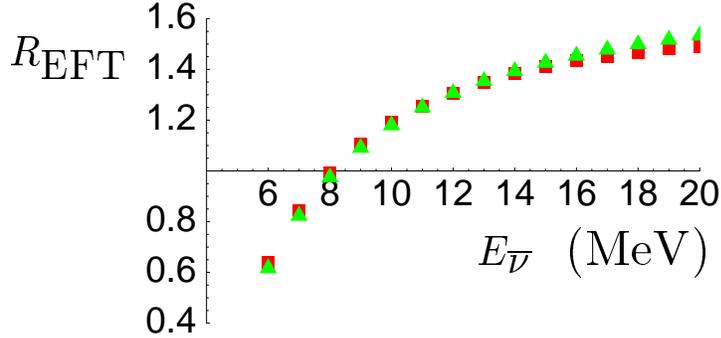}}
\caption{ 
The ratio of charged current to neutral current cross sections in
$\bar{\protect\nu} d$ scattering
versus  incident $\bar{\protect\nu}$ energy at 
NLO in EFT with $L_{1,A}=-20~{\rm fm}^3$
(boxes) and $40~{\rm fm}^3$ (triangles). 
}
\label{fig6}
\end{figure}
An important input into determining if neutrinos are changing flavor as they
move out of the sun and to the earth is the ratio of charged current to
neutral current cross sections.
Figure~\ref{fig6} shows that this ratio,
unlike the individual cross sections,
is relatively insensitive to the
counterterm $L_{1,A}$.

\section*{Low-Energy Three-Body Processes}

\begin{figure}[!t]
\epsfysize=2.5in
\centerline{\epsfbox{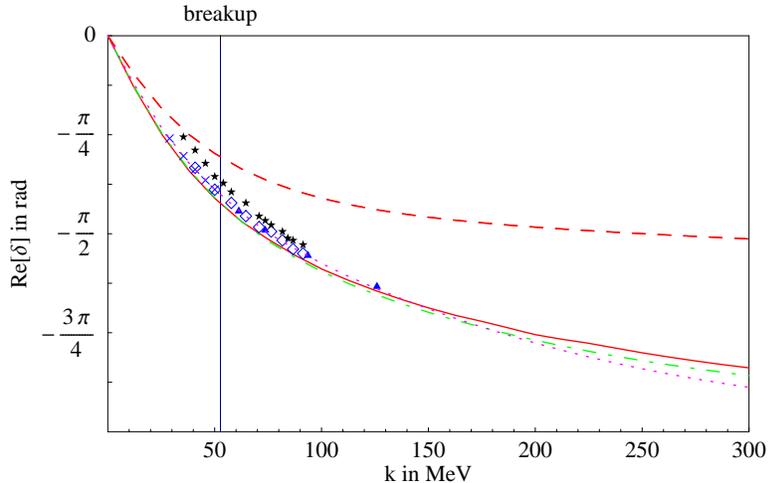}}
\caption{
${\it Re}\left(\delta\right)$ versus incident momentum for
$N d$ scattering in the  $J={3\over 2}^+$ channel.
The data are taken from the TUNL $p d$ partial wave 
analysis\protect\cite{tunl}.
The dashed, solid and dotted lines are the LO, NLO and NNLO
EFT calculations.}
\label{fig7}
\end{figure}

\begin{figure}[!t]
\epsfysize=3.3in
\centerline{\epsfbox{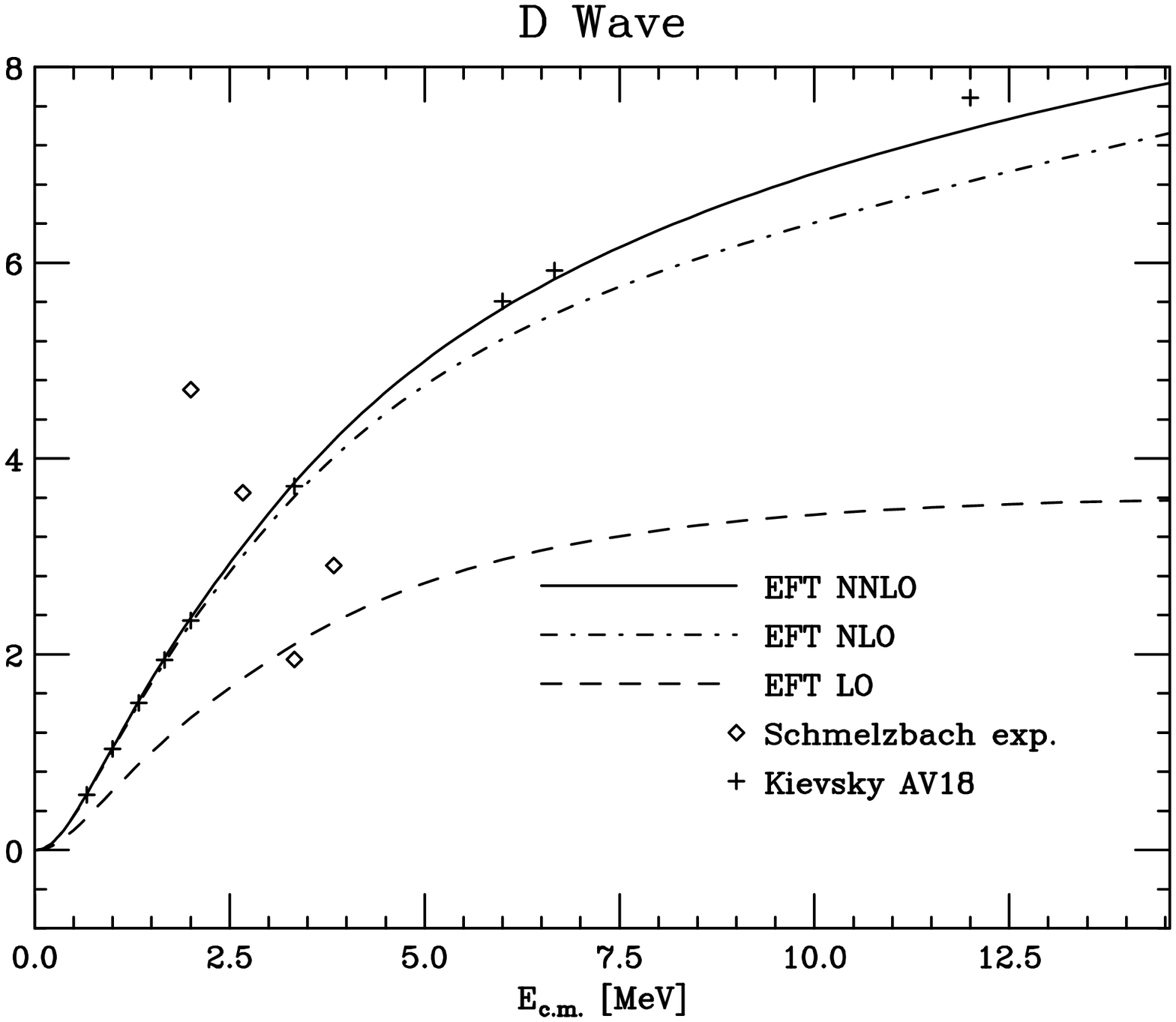}}
\vskip -0.6in
\caption{
${\it Re}\left(\delta\right)$ versus incident momentum for
$N d$ scattering in the  $L=2$ quartet channel.
The dashed, dot-dashed and solid line is LO, NLO and NNLO.
Calculations are shown as 
crosses\protect\cite{thth}, while the phase shift analysis is 
shown by open squares and diamonds~\protect\cite{thexpt}.
}
\label{fig8}
\end{figure}

Significant progress has been made in  understanding three-body systems 
with EFT\cite{threebod,triton}.
A couple of years ago, Bedaque, Hammer and 
van Kolck\footnote{I thank Paulo Bedaque, Hans-Werner Hammer and
Bira van Kolck for allowing me to reproduce their
figure.}
showed very clearly
that 
low-energy $Nd $ scattering in the $J={3\over 2}^+$ channel
could be described by EFT using contact interactions alone.
One of the more impressive results was the calculation of $a_{3\over 2}$, the 
scattering length in the quartet s-wave channel. 
Bedaque and van Kolck calculated the first three terms
to be
$a_{3\over 2} \ =\  5.01\ +\ 1.0\ +\ 0.32\ + ...~{\rm fm}$
where the ellipses denote higher order contributions,
that are estimated to be 
$\pm 0.1~{\rm fm}$.
The calculated $a_{3\over 2}=6.32\pm 0.1~{\rm fm}$ agrees very well\footnote{
Subsequent ``second generation'' potential-model calculations
agree with this result\cite{FHWPa}.}
with the 
experimental $a_{3\over 2}^{\rm expt}=6.35\pm 0.02~{\rm fm}$\cite{a32expt}.
The first term had been computed in 1957 by Skornyakov  and Ter
Martirosian\cite{SMa}
while the second term was computed in 1991 by Efimov\cite{Efi}.
The third term had not been computed before and was determined
unambiguously from the nucleon-nucleon scattering 
amplitude~\cite{threebod}.
The results of extending this analysis to non-zero energy\cite{threebod}
can be seen in Figure~\ref{fig7} and are found to agree well 
with data.
Scattering in higher partial waves has been examined by
Bedaque, Gabbiani and Grie\ss hammer\cite{BGGa} in the theory with only 
contact interactions between nucleons.
A comparison between the EFT 
calculations\footnote{I thank Paulo Bedaque, Fabrizio Gabbiani and 
Harald Grie\ss hammer for allowing me to reproduce their
figure.}, sophisticated potential
model calculations and data for one partial wave  
is shown in Figure~\ref{fig8}.

Computations in the spin-doublet channel required much more development
as local three-nucleon operators can contribute to the scattering 
amplitude\cite{triton}.
The scale-dependence of the counterterm (more commonly known as the three-body
force) is quite different from those that are familiar from 
perturbative field theory.
Bedaque, Hammer and van Kolck
showed that a single momentum independent counterterm could absorb all
cut-off dependence from  the leading operators resummed by the integral
equation that describe three-body systems.
The observed  periodicity as a function of scale
indicates that if a given calculation is performed with a given value
of the cut-off, it is possible that the three-body ``force'' vanishes,
while for a different value of the cut-off 
the three-body ``force'' may dominate.
Further, the appearance of only one three-body counterterm required to render 
the scattering amplitude scale independent also naturally
explains the {\it Phillips line}  found in potential-models.

As a last comment on the three-body work, the techniques that have
been developed in the nuclear physics setting are being applied to 
atomic systems, most notably 
scattering lengths and recombination rates in Bose condensates\cite{Bose}.

\section*{Issues at Higher Energies}

The situation regarding the applicability of 
EFT's at higher momentum, near or above
the pion mass, is much less clear.
In Weinberg's  scheme\cite{Weinberg}, pion exchange and local four-nucleon
operators contribute to nucleon-nucleon scattering at the same order
in the expansion parameter.
A great calculational and conceptual 
simplification would arise if the exchange of pions between nucleons
could be treated in perturbation theory in all partial waves,
as suggested by Kaplan, Wise and myself (KSW)\cite{KSW}.
Unfortunately,
one of the more disappointing results found during the past year is that
the EFT with perturbative pions\cite{KSW} appears not to be 
converging~\cite{FMSa}
(also, see earlier work by Cohen and Hansen\cite{cohen}).
Fleming, Mehen and Stewart~\cite{FMSa}  performed 
an analytic calculation of the   NNLO amplitude for  
nucleon-nucleon scattering in the theory with pions and KSW 
power-counting\cite{KSW}.
They found large non-analytic contributions that appear to destroy
the convergence of the series.
In contrast, several computations were performed with Weinberg's 
power-counting~\cite{Weinberg}
for the nucleon-nucleon potential
which appears to give converging amplitudes.
However, the formal inconsistency of Weinbergs power-counting remains.
Amplitudes are not renormalization (cut-off) independent at any order
in Weinberg's expansion, however, the cut-off dependence is 
found to be numerically small when 
renormalized at a typical strong interaction scale.
Therefore, at this point in time there is no formally consistent,
converging, perturbative EFT to describe nuclear interactions 
for momenta of order or higher than the mass of the pion.

\begin{figure}[!ht]
\epsfxsize=2.5in
\epsfysize=2.0in
\centerline{\epsfbox{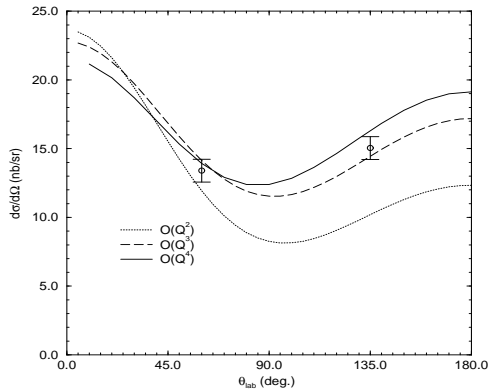}}
\caption{ ${d\sigma\over d\Omega}$ 
for $\gamma d\rightarrow\gamma d$
for incident photon energy  $E_\gamma=69~{\rm MeV}$ with 
Weinberg's power-counting\protect\cite{BMPKa}.
The dotted, dashed and solid curves are LO, NLO and NNLO.
Data is from \protect\cite{Lucas}.
}
\label{fig9}
\end{figure}

\begin{figure}[!ht]
\epsfxsize=5.0in
\centerline{\epsfbox{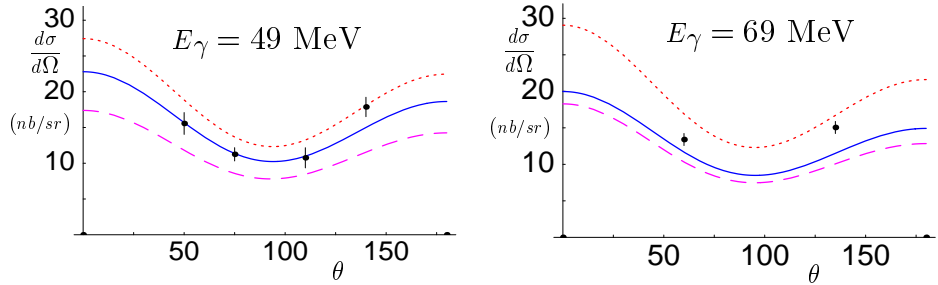}}
\caption{ ${d\sigma\over d\Omega}$ for
$\gamma d$ Compton scattering at incident photon
energies of $E_{\gamma}=49\ {\rm MeV}$  and  
$69\  {\rm MeV}$ determined
in \protect\cite{CGSSa}.
  The dashed curves are  LO.
  The solid (dotted) 
curves are NLO with (without) the graphs that
contribute to the polarizability of the nucleon.
  Data is from \protect\cite{Lucas}.
  }
\label{fig10}
\end{figure}

To give you an idea of what has been attempted at these higher energies with
both Weinberg and KSW power-counting, let me show you the results
obtained for $\gamma d\rightarrow \gamma d$, deuteron Compton scattering.
Three-orders in Weinberg's counting have been completed for 
$\gamma d\rightarrow \gamma d$ at higher energies\cite{BMPKa}.
The angular distribution of scattering photons at $E_\gamma=69~{\rm MeV}$
is shown in  Figure~\ref{fig9}\footnote{
I thank Daniel Phillips, Silas Beane and Bira van Kolck
for allowing me to reproduce their figure.}.
One can see from  Figure~\ref{fig9} that the expansion appears to be 
converging nicely to the experimental values.
Similarly, 
$\gamma d\rightarrow \gamma d$ has been computed to two orders 
in KSW power-counting\cite{CGSSa}, the results of which can be
seen in  Figure~\ref{fig10}.
Very good agreement between data 
and the parameter free-prediction at NLO
is found at  $E_\gamma=49~{\rm MeV}$.
The agreement is somewhat worse at $E_\gamma=69~{\rm MeV}$, and does 
not appear to be approaching the data in the same way that the calculation
with Weinberg's counting appears to.
It is clear that higher order calculations must be performed once the 
perturbative pions 
versus non-perturbative pions issue is understood, and further,
it is clear that more precise data is required at low-energies.
It is worth mentioning at this point that neither counting schemes, nor any
theoretical calculation that exists at present comes close to 
reproducing the recent data at $E_\gamma=95~{\rm MeV}$\cite{sask}.

There is much work still to be done in this area.

\section*{On the Road to Nuclei}

In parallel to the efforts that I have described in the two- and three-body
sectors, 
Haxton and collaborators\cite{wick} 
have been developing techniques to apply
the ideas underpinning EFT to the nuclear shell model
(efforts are ongoing by others\cite{others} but I will not discuss their
work in this talk).

\begin{figure}[!t]
\epsfysize=2.5in
\epsfxsize=3.5in
\centerline{\epsfbox{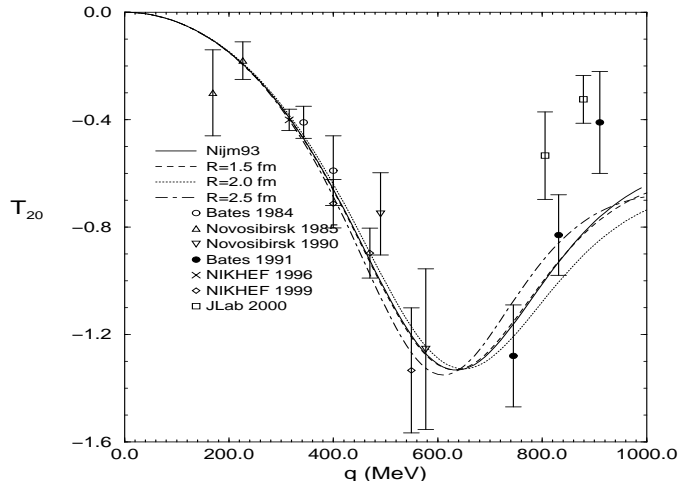}}
\caption{
$T_{20}$ computed with the Nijmegen93 potential and in the 
effective theory defined in the text for three values of the 
spatial cut-off $R$.
}
\label{fig11}
\end{figure}

Before I discuss this work I wish to show you the results of a relatively
simple but demonstrative calculation by Phillips\cite{PhilT}
(see also \cite{PCa}).
To show how the ideas of EFT can be translated into a potential-model
mode of thinking, 
Phillips compared the 
deuteron quadrupole form factor, 
presented as $T_{20}$, 
computed with the Nijmegen93 potential\cite{Nij}
with that generated by an effective potential,
$V^L_{\rm eff} (r)$ (where $L$ denotes the orbital angular momentum state)
and local quadrupole moment counterterm.
The effective potential consists of one-pion exchange at long distances
and a square-well at short-distances,
$V^L_{\rm eff} (r)=V^{\rm OPE} (r)$ for $r>R$, and  
$V^L_{\rm eff} (r)=V_{0,L}$ for $r<R$.
The values of the $V_{0,L}$ are chosen to reproduce 
the deuteron binding energy,
and low-energy nucleon-nucleon scattering
for each choice of $R$.
Therefore, the
long-distance behavior
of the ``true'' potential and effective potential
are identical.
Further, the tail of the deuteron wave-function 
produced by the ``true'' potential and effective potential are identical.
As one is brutalizing the nucleon-nucleon interaction at short-distances
while preserving nucleon-nucleon scattering, it is expected
that predictions for 
other observables, such as electromagnetic form factors, will
deviate significantly from  nature when probing distance scale comparable
to or less than $R$, for reasonable values of $R$.
In addition one expects to find that the static moments differ somewhat
from nature.
It was shown in \cite{CRSa,KSWem}
that such short-distance modifications can be compensated 
by the inclusion of gauge invariant local operators.
In the case of the deuteron quadrupole moment it is necessary
to introduce a 
four-nucleon-one-quadrupole-photon operator,
that is in no way related to the operators determining nucleon-nucleon
scattering.
These operators are induced at the chiral symmetry breaking scale, and 
must be included in any consistent calculation, 
and in fact, their omission is responsible for the discrepancy between
all sophisticated
potential model calculations of the deuteron quadrupole moment\cite{QuadPot}
and its experimental value.
Phillips choose a value of this operator to reproduce the 
observed deuteron quadrupole
moment for each value of $R$,
and then predicted $T_{20}^{\rm eff}$ 
in the effective theory, the results of which
are shown in Figure~\ref{fig11}\footnote{
I thank Daniel Phillips for allowing me to reproduce his figure.}.
It is clear from Figure~\ref{fig11} that by fixing parameters 
in the effective
theory to reproduce low-energy observables that, even with 
a nucleon-nucleon potential that has been
brutalized at short-distances,
one can essentially recover the ``true'' 
form factor over a quite impressive range of momentum transfers.
This provides a very clear demonstration that $T_{20}$ is determined
largely by the tail of the deuteron wavefunction, the OPE tail of the 
nucleon-nucleon interaction and the deuteron quadrupole moment.
For potential model calculations to accurately determine $T_{20}$,
they must first recover the deuteron quadrupole moment, and to do so 
they must include a four-nucleon quadrupole operator\cite{CRSa,KSWem}.

A very similar exploration is ongoing by Haxton and collaborators
\cite{wick}.
They are attempting to construct a 
model-space-independent shell model, and are presently focusing
on the deuteron to optimize their techniques.
The underpinnings of EFT are basis independent, and as such should be 
able to be implemented in both a plane-wave  basis 
(PWB)
or a harmonic oscillator basis (HOB).
For a continuum process, obviously the PWB is preferable, but for
a bound state it seems reasonable that a bound state basis, such as the 
HOB, will be optimal.
In an HOB basis it is natural to integrate out the levels of the HO 
step by step to allow for an easier calculation in a reduced 
model-space.
As each HO level is removed, 
the hamiltonian and coefficients of gauge invariant operators
are redefined to preserve observables.
This is a discrete analog of the renormalization group (RG)
implemented in the PWB.
\begin{figure}[!t]
\epsfxsize=4.0in
\centerline{\epsfbox{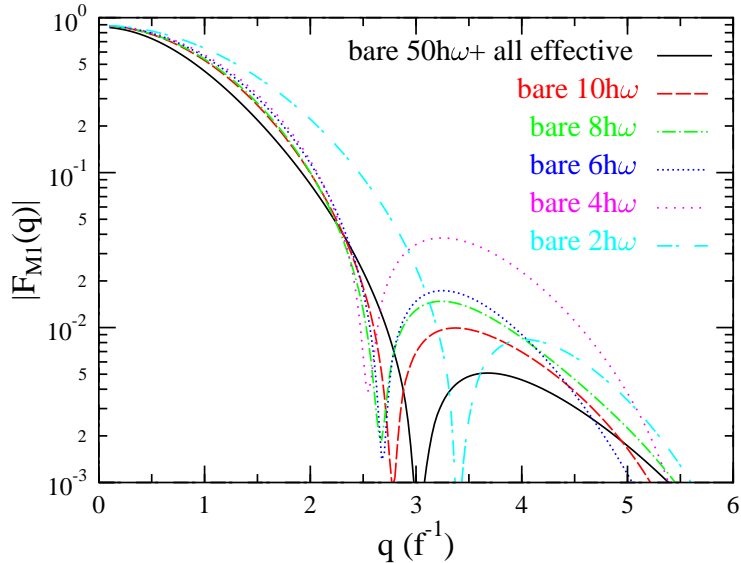}}
\vskip -2in
\caption{
The deuteron $M1$ form factor.
}
\label{fig12}
\end{figure}
In Figure~\ref{fig12}\footnote{
I thank Wick Haxton for allowing me to reproduce 
his figure.} 
the magnetic form factor of the deuteron
$F_{M1} (q)$ is shown versus momentum transfer,
where the $n=50$ calculation provides the ``true'' calculation.
The other curves correspond to the same calculation,  but with the 
insertion of the bare $M1$ operator in  each reduced model space.
The solid line, however, is not just the $n=50$ calculation but also 
the calculation from ALL reduced model-spaces when the renormalized $M1$ 
operator is inserted and NOT the bare operator.
Clearly, a discrete RG can be implemented in an HOB.

One of the 
advantages of constructing a discrete RG for the nuclear shell model
would be to 
greatly reduce the computer time required to compute matrix elements
in a nucleus with $A\gg 2$.
Presently, efforts are being made to reduce the shell-model space for the 
deuteron  down from $n=140$, which reproduces the deuteron binding energy
perfectly (by construction), down to $n=20$ or $30$ and faithfully reproduce
all deuteron observables.
Part of the effort is to include the short-range part of the 
nucleon-nucleon interaction by local operators.
If successful, this program will allow for high precision computations
of nuclear properties, with greatly increased speed.

\section*{Discussion}

I have tried to give you an overview of a 
number of important developments of the 
last year or so.
At low-energies, high precision calculations, $\sim 1\%$ have been performed
in the two-body sector, and progress is being made 
toward calculations of similar precision 
in the three-body sector.

At somewhat higher energies,
a formally consistent and  converging EFT describing nucleon-nucleon
interactions is yet to be uncovered.
Weinberg's power-counting is formally ill-defined, yet
gives numerical results that appear to be converging.
In contrast, KSW power-counting is formally well-defined, yet appears not to be
converging!
I suspect, as do others, that some sort of union between the two
power-countings may in fact be the correct one, but this is merely
speculation.

Some of the more interesting developments 
this year were made in the implementation of 
EFT ideas in nuclear many-body calculations.  
There are indications that the EFT techniques  may
provide  a means to compute properties of nuclei presently beyond
reach. However, more work is required before any conclusions can be 
drawn.

\end{document}